\documentclass[aps,prc,twocolumn,showpacs,superscriptaddress,nofootinbib,nolongbibliography,floatfix]{revtex4-2}  
\usepackage{changes} 
\usepackage[para,online,flushleft]{threeparttable}
\usepackage{graphicx}  
\usepackage{dcolumn}   
\usepackage{bm}        
\usepackage{amssymb}   
\usepackage{amssymb,amsmath,amsfonts} 
\usepackage[breaklinks=true,debug=true]{hyperref}
\usepackage{braket}    
\usepackage{booktabs}
\usepackage{multirow}
\usepackage{adjustbox}
\usepackage{gnuplottex}
\usepackage[T1]{fontenc}
\usepackage[utf8]{inputenc}
\usepackage{color}
\usepackage{dsfont}
\usepackage{tikz}
\usepackage{pgfplots}
\usepgfplotslibrary{fillbetween}
\usetikzlibrary {arrows.meta}

\usepgfplotslibrary{groupplots}
\pgfplotsset{compat=newest}
\usepgfplotslibrary{units}
\usetikzlibrary{patterns}

\begin{document}


\title{Radial excitations and their potential impact on Fermi $\beta$-decay rates}

\author{L.~Xayavong} 
\affiliation{Department of Physics, Yonsei University, Seoul 03722, South Korea}
\author{Y.~Lim}
\affiliation{Department of Physics, Yonsei University, Seoul 03722, South Korea}
\author{N.\,A.\,Smirnova}
\affiliation{LP2IB (CNRS/IN2P3-Universit\'e de Bordeaux), 33170 Gradignan cedex, France}
\author{Calvin W.\,Johnson}
\affiliation{Department of Physics San Diego State University 5500 Campanile Drive San Diego CA 92182-1233 United States}

\vskip 0.25cm  
\date{\today}

\begin{abstract} 
We investigate the contribution of radial excitations to Fermi $\beta$-decay matrix element. To this end, exact no-core shell model calculations are performed for the mirror $\beta$ decay of tritium, where full convergence can be achieved on an ordinary computer. The differences between the isospin-mixing correction values obtained in the full and in a restricted model spaces are matched to the radial overlap correction term, analogous to that required in the shell-model approach, where the configuration space is extremely limited. 
We examine this complementary correction term using a nonorthogonal harmonic-oscillator basis, generated by slightly differentiating the oscillator frequencies between the initial and final nuclei, while all desirable properties, including translational invariance, are still preserved. 
For $N_{\rm max}\le8$, we find that the radial excitation contribution is negative, with a typical magnitude of approximately 10\,\% to 20\,\% of the radial diagonal contribution. This effect becomes more pronounced as the model space increases. Therefore, the $\delta_{C2}$ values obtained in the shell model approach, where radial excitations are not explicitly included, are likely overestimated. Based on experimental $ft$ data and the corrective terms adopted in the survey by Hardy and Towner [Phys. Rev. C {\bf 102}, 045501 (2020)], we show that the incorporation of radial excitations for the superallowed $0^+\rightarrow0^+$ nuclear $\beta$ decay  tends however to worsen agreement with the Standard Model. 
\end{abstract}

\maketitle

\section{Introduction}

Isospin symmetry is an efficient tool for dealing with the nuclear many-body problem, where the system's basic constituents are protons and neutrons\,\cite{osti_4137231}. Notably, it predicts degenerate nuclear charge states within a given isospin multiplet that possess identical spin and spatial configurations. 
Additionally, it implies model-independent sum rules that has been proven to be useful for both nuclear structure and reaction studies\,\cite{FrMac1961,RevModPhys.32.567}. This symmetry is, however, imperfect because of Coulomb repulsion between protons and nuclear charge-dependent interactions\,\cite{physics5020026}. Currently, the correction due to isospin-symmetry breaking, although small, represents a limiting factor in the low-energy precision tests of the electroweak sector of the Standard Model via nuclear $\beta$ decays\,\cite{HaTo2020,ToHa2008,XaNa2018,XaNa2022,xayavong2022higherorder,NaXa2018,brodeur2023nuclearbetadecayprobe,acharya2023}. For instance, experimental precision in 15 cases of the superallowed $0^+\rightarrow0^+$ nuclear $\beta$ decay of isotriplets has reached $\mathcal{O}(10^{-4})$. These cases involve a wide variety of emitters with mass number between 10 and 74\,\cite{HaTo2020,f,PhysRevC.92.055505}, namely $^{10}$C, $^{14}$O, $^{22}$Mg, $^{26m}$Al, $^{26}$Si, $^{34}$Cl, $^{34}$Ar, $^{38m}$K, $^{38}$Ca, $^{42}$Sc, $^{42}$Ti, $^{46}$V, $^{50}$Mn, $^{54}$Co, $^{62}$Ga, $^{70}$Kr, and $^{74}$Rb. Therefore, to preserve the meaningfulness of these studies, it is mandatory to account for all contributing effects within this level of precision\,\cite{ToHa2008,PhysRevC.66.035501,xayavong2022higherorder,XaNa2018,Chien2018,PhysRevC.109.045501,SENG2023138259,PhysRevC.109.044302,PhysRevLett.130.152501,SENG2023137654,GRINYER2010236,Dam1969,PhysRevC.108.064310}. The so-called absolute $\mathcal{F}t$ value for the superallowed $0^+\rightarrow0^+$ nuclear $\beta$ decay is traditionally expressed as\,\cite{HaTo2020} 
\begin{equation}\label{eq0}
\begin{aligned}
\mathcal{F}t & =ft(1+\delta_R')(1-\delta_C+\delta_{NS})\\
& =\frac{K}{\mathcal{M}_F^2G_F^2V_{ud}^2(1+\Delta_R^V)}, 
\end{aligned}
\end{equation}
where $ft$ value is evaluated as the product of the statistical rate function ($f$) and the partial half-life ($t$)\,\cite{HaTo2020}, 
$K$ is a combination of fundamental constants, namely $K/(\hbar c)^6=2\pi^3\hbar\ln(2)/(m_ec^2)^5= 8120.27648(26)\times10^{-10}$\,GeV$^{-4}$\,s, $G_F$ is the Fermi coupling constant, precisely known from the MuLan experiment\,\cite{PhysRevD.87.052003,PhysRevD.98.030001}, and $V_{ud}$ is the top-left element of the Cabibbo–Kobayashi–Maskawa (CKM) quark-mixing matrix\,\cite{PhysRevLett.10.531,10.1143/PTP.49.652}. 
Note that $G_V^2=G_F^2V_{ud}^2$, where $G_V$ is the vector coupling constant for weak semileptonic processes. 
Further, $\mathcal{M}_F$ is the Fermi matrix element obtained within the isospin-symmetry limit and thus being model-independent, 
while other entries denote various corrections to the process. 
Radiative corrections are usually separated into three terms: $\Delta_R^V$, which is universal\,\cite{PhysRevD.100.013001,PhysRevLett.121.241804}; $\delta_R'$, which depends on the decay $Q$ value and atomic number\,\cite{PhysRevC.66.035501}; and $\delta_{NS}$, which absorbs nuclear structure effects\,\cite{TOWNER1992478,gennari2024,PhysRevC.107.035503,particles4040034,cirigliano2024}. 
Finally, $\delta_C$ is the so-called isospin-symmetry breaking correction which describes deviation of the realistic Fermi matrix element squared from its model-independent value. 

An important feature of Eq.\,\eqref{eq0} is that the constancy of the corrected $\mathcal{F}t$ value across various nuclei would serve as a direct test of the conserved vector current (CVC) hypothesis, an analogy of the conservation of electric charge current. If CVC is confirmed, the average $\mathcal{F}t$ can be used further to extract $V_{ud}$, enabling a test of the top-row unitarity of the CKM matrix. 
Deviations of the average $\mathcal{F}t$ value from a constant or breaking of the CKM matrix unitarity can be attributed to new physics effects beyond the Standard Model (presence of scalar currents, right-handed currents or else~\cite{ToHa2010}). 
The current status of these studies can be found in Ref.\,\cite{HaTo2020}. 

Radiative corrections have recently attracted a lot of attention of both particle and nuclear physicists\,\cite{gennari2024,PhysRevC.107.035503,particles4040034,PhysRevD.100.013001,PhysRevLett.121.241804,TOWNER1992478,cirigliano2024,PhysRevC.66.035501}, since new calculations lead to certain tension in the top-row CKM matrix unitarity. This in turn, encouraged nuclear theorists to put further efforts in the calculation of $\delta_C$, whose evaluation necessitates an accurate solution of the nuclear many-body problem. 
Theoretical approaches to $\delta_C$ vary widely, ranging from simple schematic models that incorporate Coulomb mixing on top of harmonic oscillator functions to fully microscopic many-body treatments, such as symmetry-restored density functional theory and the shell model. This extensive body of work has resulted in more than 30 publications in high impact journals\,\cite{ToHa2008,PhysRevC.66.035501,xayavong2022higherorder,XaNa2018,XaNa2022,AUERBACH2022122521,PhysRevC.86.054316,PhysRevC.105.065505,PhysRevC.79.064316,PhysRevLett.62.866,PhysRevC.82.065501,Calik2013,CALIK2012,Liang_2010,Barrett1998,PhysRevC.56.2542,particles4040038,PhysRevC.78.035501,PhysRevC.80.064319,PhysRevC.106.L062501,universe9050209,PhysRevC.87.054304,PhysRevC.86.034332,SAGAWA19957,ORMAND19891,KANEKO2017521,Xa2017,WILKINSON19769,WILKINSON1995421,Au2009,OrBr1995,HaTo2009,nocore}. Among all the proposed approaches, only the one based on the phenomenological shell model with realistic radial wave functions\,\cite{HaTo2020,PhysRevC.82.065501} shows remarkable agreement with the standard electroweak theory, particularly with the CVC and CKM top-row unitarity conditions. Within this consistent theoretical framework, $\delta_C$ is decomposed as 
\begin{equation}\label{eq1}
    \delta_C\approx\delta_{C1} + \delta_{C2}, 
\end{equation}
where $\delta_{C1}$ corrects for isospin mixing inside the shell-model valence space, induced by the isospin-nonconserving part of the effective Hamiltonian\,\cite{PhysRevC.87.054304,PhysRevLett.62.866,OrBr1995,ORMAND19891,KANEKO2017521}, whereas $\delta_{C2}$ accounts for contributions beyond the valence space\,\cite{ToHa2008,HaTo2009,PhysRevC.66.035501,XaNa2022,xayavong2022higherorder,XaNa2018,Xa2017}. Typically, $\delta_{C2}$ is evaluated using realistic Woods-Saxon radial wave functions, introducing a mismatch between protons and neutrons via the Coulomb and nuclear isovector potentials. Essentially, the radial wave functions are required to reproduce separation energies and charge radii, when experimental data are available\,\cite{ToHa2008,XaNa2018}. Note that several higher-order terms are excluded from Eq.\,\eqref{eq1}. These terms are generally insignificant for Fermi transitions within low isospin multiplets, including isodoublets and isotriplets\,\cite{xayavong2022higherorder}. 

Despite its remarkable success, the shell-model approach was criticized by Miller and Schwenk\,\cite{PhysRevC.80.064319} for lacking the contribution of radial excitations, which they demonstrated using a schematic model to be significant. However, no dedicated numerical calculation has yet been performed. 
While this effect might be partially captured through the use of an effective Hamiltonian and other operators, it cannot be explicitly treated within the shell model's valence spaces, which typically consist of a single oscillator shell. 
It is interesting to notice that an upper limit for the overlap $\braket{\nu1g_{9/2}|\pi0g_{7/2}}$ was experimentally determined to be 0.19\,\% in the Gamow-Teller transition of $^{207}$Hg\,\cite{BERRY2019271}. Although this overlap could also be influenced by the spin-orbit potential, as emphasized in Ref.\,\cite{xayavong2023shellmodeldescriptionisospinsymmetrybreakingcorrection}, it is primarily a manifestation of radial excitations. 
Additionally, radial excitations are the only mechanism considered by Damgaard\,\cite{Dam1969} to derive an ana\-ly\-ti\-cal formula for $\delta_C$. 
While Damgaard's model is greatly simplified, since it incorporates only nodal mixing of harmonic oscillator functions driven by a schematic one-body Coulomb potential, it effectively captures key properties of $\delta_C$, including $Z$-dependence and shell-structure effects. This result may also be regarded as evidence that radial excitations could be significant in certain scenarios. 

\section{Exact calculations using no-core shell model}\label{exact}

To verify the importance of radial excitations in a realistic calculation, 
we perform exact no-core shell model (NCSM) calculations of the total isospin-symmetry breaking correction and its individual terms in Eq.\,\eqref{eq1} to extract the radial excitation contribution. For this purpose, we consider only the mirror $\beta$ decay of tritium, $^3$H$(\beta^-)^3$He, where the computational requirements are manageable on an ordinary computer. 
Note that we use the exact $\tau_-$ operator for the calculations of the Fermi transition matrix element. NCSM calculations of $\delta_C$ for heavier nuclei are not expected to converge, even using advanced supercomputing resources. For reference, NCSM calculations for the superallowed $0^+\rightarrow0^+$ Fermi $\beta$ decay in $^{10}$C have been reported in Refs.\,\cite{nocore,BARRETT2013131}. 
The reason for this undesired behavior mainly relies on the infinite-range Coulomb interaction, which is the principal driver for isospin mixing. 
The impact of an operator's range on the convergence of its expectation values has been discussed in Ref.\,\cite{PhysRevC.71.044325}. 

In principle, the separation ansatz in Eq.\,\eqref{eq1} can be adapted to the NCSM framework, despite the lack of an appropriate strategy for ensuring asymptotic radial wave functions, as provided in the shell-model approach. By definition, the correction values deduced from the transition matrix element calculated in a finite model space specified by $N_{\rm max}$ correspond to $\delta_{C1}$ (see Appendix for the formalism). Only the value obtained from the fully converged matrix element (the value at $N_{\rm max}=\infty$) corresponds to the total correction, $\delta_C$. 
Ideally, the convergence issue may be addressed by evaluating $\delta_{C}$ through independent calculations of $\delta_{C1}$ and $\delta_{C2}$, as conducted within the shell model framework in Refs.\,\cite{xayavong2022higherorder,XaNa2018,ToHa2008}. 
However, this approach requires a realistic single-particle basis which, besides the lack of square integrability\footnote{The overlap between states with the same quantum numbers, but opposite isospin projections is not equal to unity.} and the absence of an analytical form, can lead to significant contamination from center of mass contributions, especially for light nuclei. 
To avoid these technical difficulties, we perform an independent calculation of $\delta_{C2}$ by slightly differentiating the oscillator frequencies for the initial and final nuclei, while using an isospin-invariant interaction\footnote{The reader is reminded that $\delta_{C2}$ must be evaluated with an isospin-symmetry interaction, as it corrects only for the radial mismatch of basis functions, compensating for model space truncation. The isospin admixture in many-body wave functions induced by isospin nonconserving interactions is accounted for by $\delta_{C1}$ using the conventional harmonic oscillator basis.}. 
We notice that many-body wave functions built from harmonic oscillator eigenfunctions are always factorable into intrinsic and center of mass components (provided the latter is in its lowest energy state), even when the frequencies are assumed to be isospin-dependent or freely varied for each orbits. 
Our choice of different oscillator frequencies for a mother and a daughter nuclei
meets the minimum requirement for generating radial mismatch, including radial excitations, to $\beta$ decay matrix element. In general, the fact that initial and final nuclei may have minima of their ground-state energies at different $\hbar\omega$ values is natural for the NCSM approach and the converged result is independent of this parameter. 
In spite of these benefits, we remark that our approach does not allow to improve convergence of the calculation of $\delta_C$ and therefore is applicable only to light nuclei.  
Another possible solution to construct a realistic basis, would be for example the use of an isospin dependent oscillator parameter, which on the other hand, would necessitate significant effort in calculating interaction matrix elements, demanding the generalization of the well-known Talmi-Moshinsky brackets\,\cite{EFROS2021108005,EFROS2023108852,PhysRevC.5.1534}.

\begin{figure*}[ht!]
    \centering
    \includegraphics[width=0.9\linewidth]{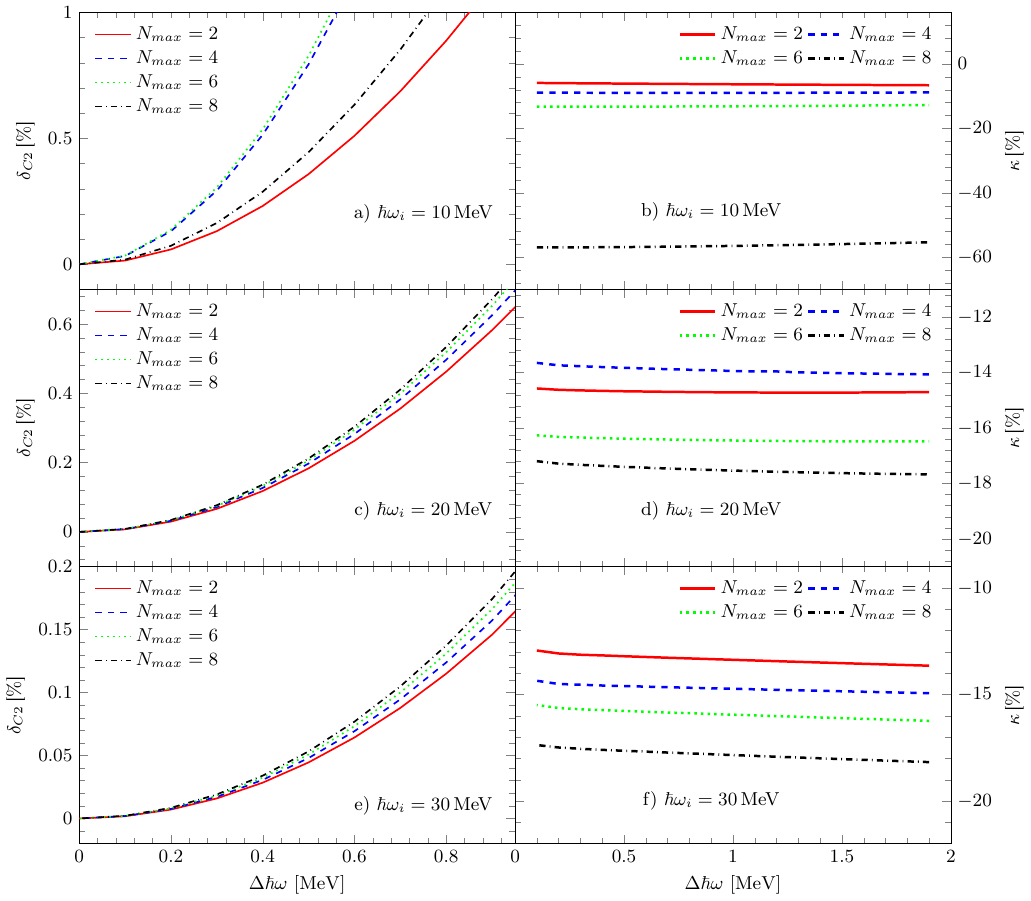}
\caption{\label{fig1}(Color online) Radial mismatch correction, $\delta_{C2}$ (left panels), and ratio $\kappa$ (right panels) for $^3$H($\beta^-$)$^3$He as a function of the oscillator frequency difference, $\Delta\hbar\omega$. The oscillator frequency for the initial state, $\hbar\omega_i$, is explicitly indicated. Note that the scales differ across panels. 
} 
\end{figure*}

We employ an interaction derived from chiral effective field theory, excluding three-body forces\footnote{As shown in Ref.\,\cite{BARRETT2013131}, three-body forces enhance the binding energy of three-nucleon systems by more than half an MeV. Noticeably, an increase in binding energy generally causes a reduction in $\delta_C$, as demonstrated in previous shell model studies\,\cite{ToHa2008,XaNa2022,XaNa2018,xayavong2022higherorder}. However, the impact of three-body forces is unlikely to be significant in our study of radial excitations, which is a higher-order effect.}. The two-body matrix elements of the Hamiltonian are generated using the NuHamil code\,\cite{Miyagi2023}, incorporating up to next-to-next-to-next-to-next-to-leading order (N4LO) terms with 500\,MeV regulator cutoff by Entem-Machleidt-Nosyk\,\cite{PhysRevC.96.024004}. The Hamiltonian is evolved via the similarity renormalization group (SRG) with a flow parameter of $\lambda_{SRG}=2$\,fm$^{-1}$. 
The isospin-symmetry interaction is obtained by switching off Coulomb repulsion and replacing the $T=1$ channel matrix elements with the average of proton-proton, neutron-neutron, and proton-neutron matrix elements. The resulting Hamiltonian matrix is diagonalized using the {\sc Bigstick} code\,\cite{johnson2018}. It is important to emphasize that when different oscillator frequencies are used for the initial and final nuclei, we must account for the overlap integrals in the evaluation of the transition matrix element. The formalism is described in detail in Appendix. Specifically, these overlap integrals reflect differences between proton and neutron radial wave functions with both identical and different numbers of nodes. When the numbers of nodes differ, this effect is referred to as radial excitations, which is the central focus of the present study. Accordingly, $\delta_{C2}$ can be broken as 
\begin{equation}\label{nn}
    \delta_{C2} = \delta_{C2}^{sm}+\delta_{C2}^{re}=\delta_{C2}^{sm}(1+\kappa), 
\end{equation}
where $\delta_{C2}^{sm}$ and $\delta_{C2}^{re}$ represent the contributions arising from the radial diagonal (obtained in the shell model approach) and radial excitation mechanisms, respectively, while $\kappa$ is given by the ratio $\delta_{C2}^{re}/\delta_{C2}^{sm}$. 
Although the asymptotic wave functions still retain a Gaussian form and the resulting correction terms might be unrealistic, the ratio $\kappa$ is expected to be reliable, showing no substantial dependence on the mean field potential model. 
The results for $\delta_{C2}$ and $\kappa$ as a function of the oscillator frequency difference, $\Delta\hbar\omega=\hbar\omega_i-\hbar\omega_f$, are shown in Fig.\,\ref{fig1}. 
We remark that, by construction, $\delta_{C2}$ is symmetric with respect to $\Delta\hbar\omega=0$. However, since neutrons in the initial nucleus are more tightly bound than protons in the final nucleus, it is generally more natural to set $\Delta\hbar\omega$ to be slightly larger than 0. Note that we consider only small $\Delta\hbar\omega$; otherwise, Eq.\,\eqref{eq1}, which will be employed in the following, would no longer be justified, as the higher-order terms would become significant. It is observed that $\delta_{C2}$ follows a quadratic trend, which can be understood from its Taylor expansion, 
\begin{equation}\label{eqx0}
\delta_{C2}(\Delta\hbar\omega) = \delta_{C2}'(0)\Delta\hbar\omega + \frac{\delta_{C2}''(0)}{2!}\Delta\hbar\omega^2+\mathcal{O}(\Delta\hbar\omega^3), 
\end{equation}
with $\delta_{C2}(0)=0$. Eq.\,\eqref{eqx0} is generally $N_{\rm max}$ dependent. 

It is interesting to emphasize that the isospin admixture spreads the model-independent Fermi matrix element [obtained under SU(2) symmetry] across all possible channels\footnote{The total Fermi strength is conserved, and is given by the sum of the allowed and isospin forbidden transitions.}, including isospin forbidden ones. This implies that $\delta_{C}$ applied to a particular channel must always be positive, resulting in a reduction in the magnitude of the actual Fermi matrix element. Previous shell model studies suggest that this behavior also holds for the individual correction terms, including $\delta_{C1}$ and $\delta_{C2}^{sm}$. Only the higher-order terms, such as $\delta_{C2}^{re}$, and some other terms discussed in Ref.\,\cite{xayavong2022higherorder}, may be negative.  


\begin{figure}
    \centering
    \includegraphics[width=0.95\linewidth]{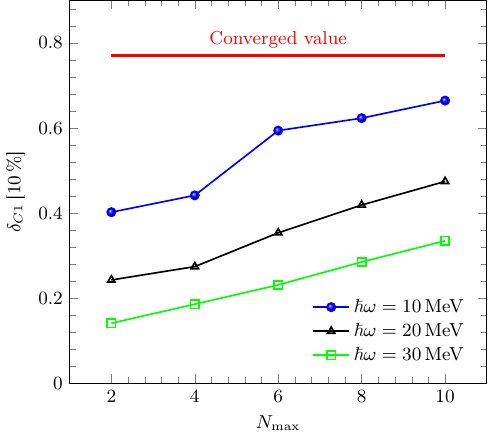}
\caption{\label{fig2}(Color online) Illustration of the isospin-mixing ($\delta_{C1}$) contribution as a function of the model space size, $N_{\rm max}$, for $^3$H$(\beta^-)^3$He (mirror $\beta$ decay). 
The correction values $\delta c_1$ are red scaled by a factor of 10. The difference between the calculated $\delta_{C1}$ value at a finite model space and its converged value defines the radial mismatch ($\delta_{C2}$) contributions. By definition, $\delta_{C2}=0$ at $N_{\rm max}=\infty$. 
See text for discussion of convergence and converged value.
}
\end{figure}
As seen in the right panels of Fig.\,\ref{fig1}, the radial excitation contribution is consistently negative. These results are in agreement with those of Miller and Schwenk\,\cite{PhysRevC.80.064319}. 
We remark that the ratio $\kappa$ varies very slowly with $\Delta\hbar\omega$, in contrast to $\delta_{C2}$, whose behavior follows Eq.\,\eqref{eqx0}, as evident from the left panels of Fig.\,\ref{fig1}. However, $\kappa$ shows a strong dependence on $N_{\rm max}$. For instance, at $\hbar\omega_i=10$\,MeV, $\kappa$ remains nearly constant for $N_{\rm max}\le6$, but jumps by almost an order of magnitude (with a negative sign) at $N_{\rm max}=8$. This dependence weakens as $\hbar\omega_i$ increases. 
At $\hbar\omega_i=30$\,MeV, the $\kappa$ values obtained for model spaces of $2\hbar\omega$, $4\hbar\omega$, $6\hbar\omega$, and $8\hbar\omega$, are approximately $-13$\,\%, $ -15$\,\%, $-16$\,\%, and $-18$\,\%, respectively. This observation is due to the fact that larger model spaces include more orbits with different nodes, leading to stronger radial excitation contribution. 
Both $\delta_{C2}^{sm}$ and $\delta_{C2}^{re}$ must vanish at $\Delta\hbar\omega=0$ where the orthogonality is fully restored. Since the magnitude of $\delta_{C2}^{re}$ is always smaller than that of $\delta_{C2}^{sm}$, $\kappa$ must also be 0 at $\Delta\hbar\omega=0$ (our results for $\kappa$ do not include $\Delta\hbar\omega=0$ because of numerical instability). 
In the present case, we have $\kappa=0$ for the $0\hbar\omega$ model space because no nodal mixing occurs in the lowest energy configurations of the initial and final nuclei. Nodal mixing in the lowest energy configuration starts to appear only from $sd$-shell nuclei. Therefore, $p$- or $s$-shell nuclei, such as tritium, are free from nodal mixing with core orbits. 
In general, $\delta_{C2}$ decreases with increasing $\hbar\omega_i$ due to the binding energy effect\footnote{As the potential depth increases, the radial mismatch becomes less sensitive to the isovector counterpart.}, while its variation with the frequency difference follows the opposite trend as described earlier. 
Thus so far, the key properties of $\delta_{C2}$ and its individual components, including radial excitations, are seen to be well described and are consistent with other studies. The ratio $\kappa$ can already be approximately estimated, at least for its lower limit, as it is only weakly dependent on the oscillator parameters (except for small $\hbar\omega_i$, e.g., $\hbar\omega_i\le10$\,MeV, where the dependence is somewhat stronger). 
However, since full convergence is achievable for the decay of tritium, we can constrain $\delta_{C2}$ by using the values extracted as the difference between $\delta_C$ and $\delta_{C1}$, according to Eq.\,\eqref{eq1}. 

Our calculation of $\delta_{C1}$ uses the conventional orthogonal harmonic oscillator basis and isospin-nonconserving interaction. This interaction is similar to the one described earlier, but without enforcing an isospin-symmetry restoration. Our calculation achieves full convergence at $N_{\rm max}\approx30$, where the conditions $\delta_{C}=\delta_{C1}$ and $\delta_{C2}=0$ are satisfied. We obtain $\delta_C=0.077$\,\%. 
Our results for $\delta_{C1}$ for various oscillator frequencies are displayed as a function of $N_{\rm max}$ in Fig.\,\ref{fig2}. 
The numerical values of $\delta_{C1}$ and the corresponding extracted $\delta_{C2}$ are listed in the third and fourth columns of Table\,\ref{tab1}, respectively. 
We observe that at smaller $\hbar \omega _i$ values, $\delta_{C1}$ approaches its final (converged) value more rapidly as $N_{\rm max}$ increases. This trend is expected, since smaller $\hbar\omega_i$ values better simulate the weakly bound nature of light nuclei\footnote{The Gaussian tail of harmonic oscillator functions can be well approximated with an exponential at the limit when the frequency is close to 0}. For the considered systems with 3 nucleons, the optimal absolute $\hbar\omega_i$ value is found to be around 10 to 13\,MeV. 
\begin{center}
\begin{table}[ht!]
\caption{Results for $\Delta\hbar\omega$ and $\kappa$ constrained by the extracted $\delta_{C2}$ values. The units of oscillator frequencies and their frequency differences are MeV, whereas those of $\delta_{C1}$, $\delta_{C2}$ and $\kappa$ are \%. The values for $\delta_{C1}$ are obtained from exact NCSM calculations, while the other values are extracted as discussed in the main text.}
\label{figt}
\begin{ruledtabular}
\begin{tabular}{ c|c|c|c|c|c } 
$\hbar\omega_i$ & $N_{\rm max}$ & $\delta_{C1}$ &	$\delta_{C2}$ &	$\Delta\hbar\omega$ &	$\kappa$ \\
 \hline
\multirow{4}{*}{10} & 2 & 0.402	& 0.037 &	0.51 &	-5.97 \\
&4 & 0.442 &	0.033	& 0.32 &	-8.93 \\
&6 & 0.594 &	0.018 &	0.23 &	-13.24 \\
&8 & 0.623 &	0.015	& 0.28 &	-56.92 \\
\hline 
\multirow{4}{*}{20} & 2 & 0.24	& 0.053 &	0.85 &	-14.70 \\
&4 & 0.274 &	0.049	& 0.80 &	-14.00 \\
&6 & 0.354 &	0.042 &	0.72 &	-16.42 \\
&8 & 0.419 &	0.035	& 0.64 &	-17.45 \\
\hline		
\multirow{4}{*}{30} & 2 & 0.141	& 0.064 &	0.60 &	-13.23 \\
&4 & 0.186 &	0.059	& 0.55 &	-14.62 \\
&6 & 0.231 &	0.055 &	0.52 &	-15.78 \\
&8 & 0.285 &	0.049	& 0.48 &	-17.63 \\
\end{tabular}
\end{ruledtabular}
\label{tab1}
\end{table}
\end{center} 

By projecting the extracted $\delta_{C2}$ values onto the results illustrated in Fig.\,\ref{fig1}, we deduced the $\Delta\hbar\omega$ values for a given model space and $\hbar\omega_i$ as listed in the fifth column of Table\,\ref{tab1}. Generally, different values of $\Delta\hbar\omega$ are required for different $\hbar\omega_i$ and $N_{\rm max}$ to produce the correct $\delta_{C2}$ that bridges the gap between $\delta_{C1}$ and $\delta_{C}$. For the considered ranges of $\hbar\omega_i$ and $N_{\rm max}$, we find the optimal $\Delta\hbar\omega$ varies between 0.23 and 0.85\,MeV. 
At $\hbar\omega_i=10$\,MeV, the deduced $\kappa$ substantially changes between the $2\hbar\omega$ and the $8\hbar\omega$ model spaces, from $-5.973$\,\% to $-56.921$\,\%. Whereas, at $\hbar\omega_i=20$ and 30\,MeV, it varies only from $-13.234$\,\% and $-17.628$\,\%. 
Based on these results, it is safe to conclude that, on average, the radial excitation contribution is stronger than $-10$\,\%. 
Despite the considerable parameter and space dependence, these results suggest that $\delta_{C2}^{re}$ could be significant for the superallowed $0^+\rightarrow0^+$ nuclear $\beta$ decay of isotriplets, especially in medium and heavy nuclei, where multiple oscillator shells are simultaneously occupied. 

\section{Standard-Model tests of radial excitations} 

In order to test the potential impact of this effect on the shell model calculations of $\delta_C$, we extract $\delta_{C2}^{re}$ by reversing the usual procedure used in the superallowed $0^+\rightarrow0^+$ nuclear 
$\beta$ decay studies. Under the CKM's top-row unitarity and using the $V_{us}$ value recommended by the Particle Data Group in 2018\,\cite{PhysRevD.98.030001}, we deduce $|V_{ud}^{st}|^2=0.94969(22)$. Here the label $st$ indicates that $V_{ud}^{st}$ is evaluated under the Standard Model framework. In this manner, the corrected $\mathcal{F}t$ is uniquely obtained via 
\begin{equation}
    \mathcal{F}t^{st}=\frac{K}{2G_F^2(1+\Delta_R^V)|V_{ud}^{st}|^2} = \frac{2912.95\pm0.54}{|V_{ud}^{st}|^2}, 
\end{equation} 
where the value for $\Delta_R^V$ is taken from Ref\,\cite{HaTo2020}. 
This Standard Model constraint yields $\mathcal{F}t^{st}=3067.26(91)$\,s. 

Furthermore, we decompose $\mathcal{F}t$ into two parts according to Eq.\,\eqref{nn} as 
\begin{equation}\label{fac}
    \mathcal{F}t(\kappa)=\mathcal{F}t^{sm}-ft(1+\delta_R')\kappa\delta_{C2}^{sm}, 
\end{equation}
where $\mathcal{F}t^{sm}$ represents the conventional corrected $\mathcal{F}t$ evaluated at the limit of $\delta_{C2}^{re}=0$. Whereas the second term in Eq.\,\eqref{fac} is induced by $\delta_{C2}^{re}$. For simplicity, $\kappa
$ is assumed to be nucleus-independent. Under these assumptions, we can effectively determine $\kappa$ by adjusting it such that the $\chi^2/\nu$ defined below is minimized 
\begin{equation}\label{chi}
    \left[\chi^2/\nu\right]^{2c} = \frac{1}{\nu}\sum_{i=1}^{N}\frac{\left[\mathcal{F}t_i(\kappa)-\mathcal{F}t^{st}\right]^2}{\sigma_i^2}, 
\end{equation}
where $\nu=N-1$, standing for the number of degree of freedoms, with $N$ being the sample size (number of transitions). As mentioned in the introduction, the uniqueness or constancy of the corrected $\mathcal{F}t$ is an implication of CVC. Therefore, the $\chi^2/\nu$ defined in Eq.\,\eqref{chi} provides a sensitive measure of consistency with CVC. The index $2c$ in Eq.\,\eqref{chi} indicates that two Standard-Model constraints are employed: CVC and CKM's top-row unitarity condition. Note that we consider only the best 15 cases selected in Ref.\,\cite{HaTo2020} with experimental precision of 0.1\,\% or better. The variant $\sigma_i$ is taken as the quadratic sum of the uncertainties on $\mathcal{F}t_i^{sm}$ and $\mathcal{F}t_i^{st}$. The data for $\mathcal{F}t_i^{sm}$, $ft_i$, $\delta_{Ri}'$, and $\delta_{C2i}^{sm}$ are taken from Ref.\,\cite{HaTo2020}. 

\begin{figure}[ht!]
    \centering
    \includegraphics[width=0.95\linewidth]{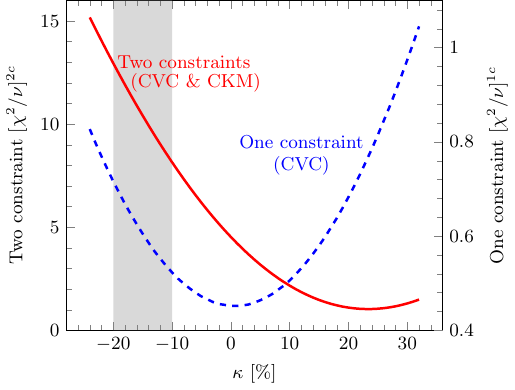}
\caption{\label{fig3}(Color online) Illustration of $[\chi^2/\nu]^{1c}$ and $[\chi^2/\nu]^{2c}$ defined in Eq.\,\eqref{chi} and \eqref{chinew}, respectively, as function of $\kappa$. 
The gray band represents the typical range of $\kappa$ obtained in the present NCSM calculations.
} 
\end{figure}

Fig.\,\ref{fig3} (the solid red curve) shows that the shape of the $[\chi^2/\nu]^{2c}$ curve is asymmetric with respect to $\kappa=0$, with positive $\kappa$ values producing better consistency. This is not surprising as the used set of $\mathcal{F}t_i$ values is known to underestimate the CKM's top-row unitarity condition by more than two standard deviations\,\cite{HaTo2020}. In this situation, one way to improve the agreement with the Standard Model is to reduce the average of $\mathcal{F}t_i$. Therefore, a negative $\kappa$ as predicted by Miller and Schwenk and also indicated by our calculations in Sec.\,\ref{exact}, which results in an increase of $\mathcal{F}t_i$ [see Eq.\,\eqref{fac}], does not align with this requirement. In this test, the Standard-Model optimized $[\chi^2/\nu]^{2c}$ corresponds to $\kappa\approx25\,\%$, which is clearly opposite to the expectation from theories. 

Another Standard Model test is conducted based solely on CVC, using the following formula
\begin{equation}\label{chinew}
    \left[\chi^2/\nu\right]^{1c} = \frac{1}{\nu}\sum_{i=1}^{N}\frac{\left[\mathcal{F}t_i(\kappa)-\overline{\mathcal{F}t}(\kappa) \right]^2}{\sigma_i^2}, 
\end{equation}
where the index $1c$ indicates that only one Standard-Model constraint is employed. Unlike the previous test, which compares the individual corrected $\mathcal{F}t_i$ values with the unique value extracted via the CKM's sum rule, Eq.\,\eqref{chinew} utilizes the sample mean (weighted average) as the reference value, evaluated according to the actual value of $\kappa$. Therefore, the goal of this test is ultimately to find the optimal $\kappa$ value by adjusting it to minimize the intrinsic scatter in the data sample, as implied by CVC. It is seen from Fig.\,\ref{fig3} (the dashed blue curve) that the minimum value of $\left[\chi^2/\nu\right]^{1c}$ corresponds to $\kappa\approx0$, which contradicts both the result of our first test and the theoretical expectation (as per Miller and Schwenk's criticism). Nevertheless, this CVC test is considerably less sensitive to $\kappa$ compared to the first test. Specifically, $\left[\chi^2/\nu\right]^{1c}$ remains below 1 for $-30\,\%\le\kappa\le30\,\%$. Additionally, we have verified that the corrected $\mathcal{F}t$ remain with the error bars despite considerable variations in $\kappa$, i.e., between $-20$\,\% and 20\,\%. This reduced sensitivity arises from allowing the reference value in the test to vary with $\kappa$, leading to a greater flexibility than using a fixed reference value. 
This inconsistency between these tests and our NCSM calculations may arise from inaccuracies in experimental or theoretical inputs, particularly the leading nuclear structure dependent corrections. Otherwise, it could indicate a violation of CVC or an incompleteness in the CKM sum rules. 

\section{Conclusion}

We present a comprehensive study of radial excitations and its influence on nuclear matrix elements of $\beta$ decays. Using no-core shell model with an isospin-dependent harmonic oscillator basis, we show that the radial excitation contribution is negative and can be significant in first-principle calculations within a large model space, in full agreement with the prediction
of Miller and Schwenk. 
Although it is not entirely clear whether this constitutes a significant defect of the phenomenological shell model approach, as the effect may be implicitly captured through adjustments of the model's parameters, it is explicitly excluded in the evaluation of the transition matrix elements in valence spaces. 
By contrast, our analysis, based on both CVC and CKM's top-row unitarity, suggests a significant positive radial excitation contribution. Conversely, this effect is disfavored under CVC when the constraint from the CKM sum rule is omitted. 
In addition to the relevant physics discussed, the present study also serves as a proof of principle for ab initio nuclear structure calculations using a nonorthogonal harmonic oscillator basis. More flexible scenarios, such as unequal oscillator frequencies for protons and neutrons or even state dependent frequencies, could be implemented without losing the desired properties obtained with the conventional case. Although the asymptotics of these generalized oscillator functions remain unrealistic, the relaxation could provide sufficient flexibility to capture essential effects and potentially improve computational efficiency. Extending the approach in this direction is left for future work, as it requires significant effort in transforming the interaction matrix elements from arbitrary coordinates to intrinsic system of interacting nucleons. 
For further investigation, the contribution of radial excitations could be addressed in a valence space by employing realistic nuclear interactions and constructing effective transition operators. 
\begin{acknowledgments} 
L.~Xayavong would like to thank Prof. A.~M.~Shirokov for useful discussions on various aspects of the no-core shell model. 
The technical support and provision of the \texttt{NuHamil} code by T.~Miyagi are gratefully acknowledged.
L.~Xayavong and Y.~Lim are supported supported by the National
Research Foundation of Korea(NRF) grant funded by the
Korea government(MSIT)(No. RS-2024-00457037) and
by Global - Learning \& Academic research institution for Master's·PhD students, 
and Postdocs(LAMP) Program of the National Research Foundation of Korea(NRF) grant funded by the Ministry of Education(No.  RS-2024-00442483).
Y. Lim is also supported by the Yonsei University Research Fund of 2024-22-0121. 
N.~A.~Smirnova acknowledges the financial support of CNRS/IN2P3, France, via ABI-CONFI Master project. 
The contribution of C.~W.~Johnson is based upon work supported by the U.S. Department of Energy, Office of Science, Office of Nuclear Physics, 
under Award Number  DE-FG02-03ER41272.

\end{acknowledgments}

%

\newpage

\appendix*
\section{Transition matrix elements in nonorthogonal basis}

In second quantization, a one-body operator such as those responsible for Fermi and Gamow-Teller transitions is written as 
\begin{equation}\label{op}
O_x = \sum_{\alpha\beta}^{N_{sp}} \braket{\alpha|O_x|\beta}c_\alpha^\dagger c_\beta, 
\end{equation}
with $x$ standing for Fermi or Gamow-Teller. $N_{sp}$ represents the number of single-particle states available within the chosen model space. It should be noted that the basis used in Eq.\,\eqref{op} is assumed to be orthogonal such that the following conditions are fulfilled: $\{c_\alpha^\dagger, c_\beta^\dagger\}=\{c_\alpha, c_\beta\}=0$ and $\{c_\alpha, c_\beta^\dagger\}=\delta_{\alpha\beta}$. The Greek letters $\alpha$ and $\beta$ denote the set of spherical quantum numbers characterizing single-particle states, $nljm$. 

The reduced matrix element of $O_x$ between initial $\ket{\Psi_i}$ and final $\ket{\Psi_f}$ states is given by 
\begin{equation}\label{reduce}
    M_x = \sum_{k_\alpha k_\beta}^{N_{sp}} \braket{k_\alpha||O_x||k_\beta} \braket{\Psi_f||[c_{k_\alpha}^\dagger \otimes \tilde{c}_{k_\beta}]^{(\lambda_x)}||\Psi_i}, 
\end{equation}
with $\lambda_x$ specifying the tensor rank of $O_x$ which is 0 for Fermi and 1 for Gamow-Teller. Here $k$ denotes the set of spherical quantum numbers, similar to $\alpha$ and $\beta$ described above, excluding $m$. The expressions for $\braket{k_\alpha||O_x||k_\beta}$ is operator specific and can be found in standard nuclear structure textbooks. 

The isospin-symmetry breaking correction $\delta_C$ is defined as 
\begin{equation}
    \displaystyle\delta_C=\left(1-\frac{M_x^2}{\mathcal{M}_x^2}\right), 
\end{equation}
where $\mathcal{M}_x$ represents the transition matrix element at isospin-symmetry limit, evaluated in the same model space as $M_x$. For Fermi transition, the isospin-symmetry matrix element is analytically known: $\mathcal{M}_F^2=T(T+1)-T_{zi}T_{zf}$. Contrary, $\mathcal{M}_{GT}$ remains model-dependent. 

We are interested in the scenario where the many-body states $\ket{\Psi_i}$ and $\ket{\Psi_f}$ are constructed on different harmonic oscillator bases 
\begin{equation}
    \ket{\Psi_i} = \sum_y^{N_{sd}} \xi_i^y \ket{\Phi_i^y} = \sum_y^{N_{sd}} \xi_i^y [\prod_\gamma^A a_{\gamma}^\dagger]_y\ket{-}, 
\end{equation}
and 
\begin{equation}
    \ket{\Psi_f} = \sum_x^{N_{sd}} \xi_f^x \ket{\Phi_f^x} = \sum_x^{N_{sd}} \xi_f^x [\prod_\gamma^A b_\lambda^\dagger]_x\ket{-}, 
\end{equation}
where $\ket{-}$ and $N_{sd}$ denotes the vacuum and the number of Slater determinants spanning the many-body basis, respectively. The coefficients $\xi_i^y$ and $\xi_f^x$ represent the wave function amplitudes that can be obtained through the diagonalization of the Hamiltonian matrix. For simplicity, we assume that the bases of the initial and final states are intrinsically orthogonal but nucleus-dependent. These bases can be formed with harmonic oscillator eigenstates using different oscillator frequencies for the initial and final nuclei. 

With this ansatz for the many-body wave functions, the one-body probability density matrix can be expressed as
\begin{equation}\label{z}
\begin{array}{ll}
    \braket{\Psi_f|c_\alpha^\dagger c_\beta|\Psi_i}&\displaystyle=\sum_{xy}^{N_{sd}} \xi_f^{x*} \xi_i^y \braket{\Phi_f^x|c_\alpha^\dagger c_\beta|\Phi_i^y} \\[0.1in]
    &=\displaystyle \sum_{xy}^{N_{sd}} \xi_f^{x*} \xi_i^y\braket{-|b_{\lambda_A}...b_{\lambda_1} c_\alpha^\dagger c_\beta a_{\gamma_1}^\dagger...a_{\gamma_A}^\dagger|-}_{xy}. 
\end{array}
\end{equation} 

To evaluate the matrix elements in Eq.\,\eqref{z}, we perform basis transformation as 
\begin{equation}\label{x}
c_\beta=\sum_\lambda^{N_{sp}}\tilde{\delta}_{\beta\lambda}^* b_\lambda, 
\end{equation}
where the overlap matrix is given by  $\tilde{\delta}_{\beta\lambda}^*=\Omega_{n_\beta n_\lambda} \delta_{l_\beta l_\lambda} \delta_{j_\beta j_\lambda} \delta_{m_\beta m_\lambda}$
with the radial component of 
\begin{equation}
    \Omega_{nn'}=\int_0^\infty R_n(r)R_{n'}(r)r^2dr, 
\end{equation}
it is interesting to note that for the present case, $\tilde{\delta}_{nn'}$ is invariant by exchanging $n$ and $n'$. Additionally, $\Omega_{nn'}$ is slightly less than 1 when $n=n'$ and slightly greater than 0 when $n\ne n'$ given that $\Delta\hbar\omega\ne0$. 

Similarly,  
\begin{equation}\label{y}
c_\alpha^\dagger=\sum_\gamma^{N_{sp}}\tilde{\delta}_{\alpha\gamma}a_\gamma^\dagger, 
\end{equation}
where $\tilde{\delta}_{\alpha\gamma}=\Omega_{n_\alpha n_\gamma} \delta_{l_\alpha l_\gamma} \delta_{j_\alpha j_\gamma} \delta_{m_\alpha m_\gamma}$. 

Substituting Eqs.\,\eqref{x} and \eqref{y} into Eq.\,\eqref{z}, we arrive at 
\begin{widetext}
\begin{equation}\label{t}
\begin{array}{ll}
    \braket{\Psi_f|c_\alpha^\dagger c_\beta|\Psi_i}&=\displaystyle \sum_{xy}^{N_{sd}} \xi_f^{x*} \xi_i^y \sum_{\lambda\gamma}^{N_{sp}} \tilde{\delta}_{\alpha\gamma} 
    \tilde{\delta}_{\beta\lambda} 
    \braket{-|b_{\lambda_A}...b_{\lambda_1}b_{\lambda}^\dagger a_{\gamma}a_{\gamma_1}^\dagger...a_{\gamma_A}^\dagger|-}_{xy}. 
\end{array}
\end{equation} 
\end{widetext} 

Obviously, the choice of the common basis $\{c_\alpha^\dagger\}$ is trivial. It can be chosen to be equal to $\{a_\gamma^\dagger\}$ or $\{b_\lambda^\dagger\}$ such that one summation in Eq.\,\eqref{t} disappears. 

The matrix element on the right-hand side of Eq.\,\eqref{t} can be written in a compact form as 
\begin{equation}
\braket{-|b_{\lambda_A}...b_{\lambda_1}b_\lambda^\dagger a_\gamma a_{\gamma_1}^\dagger...a_{\gamma_A}^\dagger|-} 
= \displaystyle \tilde{\delta}_{\lambda\gamma} \prod_k^{\lambda_k\gamma_k\ne\lambda\gamma} \tilde{\delta}_{\lambda_k\gamma_k}, 
\end{equation}
if $a_\gamma \ket{\Phi_i^x}=1$ and $\bra{\Phi_f^y}b_\lambda^\dagger=1$. Otherwise, it vanishes. 

Finally, the reduced matrix element $\braket{\Psi_f||[c_\alpha^\dagger \otimes\tilde{c}_\beta]^{(\lambda_x)}||\Psi_i}$ in Eq.\,\eqref{reduce} can be easily obtained from Eq.\,\eqref{t} by using the Wigner–Eckart theorem. 
In this study, our numerical implementation of nonorthogonal basis is based on the so-called odometer method. A more efficient method would be required for heavier nuclei. More details on this computational aspect can be found in Refs.\,\cite{UTSUNO2013102,TOGASHI20141711,PhysRevA.101.012105,scemama2013,PhysRevC.105.054314}.




\end{document}